\documentstyle[preprint,aps,floats]{revtex}
\tightenlines

\input psfig.tex 

\begin{document}

\title{\bf Quantization of the massless minimally coupled scalar field
 and the dS/CFT correspondence}
\author{Andrew J. 
Tolley\thanks{Email:A.J.Tolley@damtp.cam.ac.uk}, 
Neil Turok\thanks{Email:N.G.Turok@damtp.cam.ac.uk}}
\address{
 DAMTP, Centre for Mathematical Sciences, Cambridge, CB3 0WA, U.K.}
\date{\today}

\maketitle

\begin{abstract}
We consider the quantization of the massless minimally coupled scalar
field in de Sitter spacetime. The no-boundary Euclidean prescription
naturally picks out the de Sitter invariant vacuum state of Kirsten
and Garriga. We extend the dS/CFT correspondence to this case
which allows us to interpret the massless field in terms of a Euclidean
CFT. The extension is non-trivial and requires careful treatment of
the zero mode.
\end{abstract}
\vskip .2in

\section{Introduction}

De Sitter spacetime is one of the simplest and most interesting spacetimes
allowed by general relativity. If current indications from supernovae,
cosmic microwave background and other astronomical data have been interpreted
correctly, it seems that the Universe we live in will in fact be asymptotically
de Sitter in the infinite future. 
De Sitter spacetime also plays a central role in the theory of inflation,
where an approximately de Sitter spacetime, with
exponential expansion, is employed to
solve the cosmological flatness and horizon puzzles. 

The quantum mechanics of de Sitter spacetime, and matter fields within it,
is also of considerable interest. During inflation, massless quantum fields
acquire fluctuations on macroscopically large scales, leading to
an attractive theory of the origin of structure in the Universe. 
Indeed this setup provides one of the very few direct observational
probes of quantum gravity, since the gravitational waves generated
during such a period of early inflation may be directly observable 
today through the polarization signature they produce 
on the cosmic microwave sky.

In this article we consider the quantization of the massless minimally
coupled scalar field in de Sitter spacetime and its relation to a
conformal field living at the boundary. The massless case must be treated
separately from the massive case due to the presence of the symmetry
$\phi \rightarrow \phi +q$ where $q$ is a space-time constant. 
This symmetry is relevant for a massless
minimally coupled scalar field on any spacetime with compact spatial
sections, whereas for spacetimes (such as Minkowski spacetime) 
with infinite spatial sections, super-selection rules 
render the symmetry innocuous.
As shown by Allen and Folacci \cite{Allen} the presence of
the zero mode implies there is no de Sitter invariant Fock vacuum for
the massless field. Nevertheless there is still a de Sitter invariant vacuum
as shown by Kirsten and Garriga \cite{Garriga}. We will show that it
is this latter vacuum state that is naturally picked out by the no boundary
 Euclidean
prescription. 

Recently Strominger \cite{Strominger} has given
arguments showing that a scalar field of mass $0 < m^2 < (\frac{d}{2}H)^2$
in $dS^{d+1}$
can be related to a conformal field theory living on an $S^d$ which
may be identified as the boundary (timelike infinity) of de Sitter
spacetime\cite{Bros}. This proposal was formulated in analogy with the AdS/CFT
correspondence and arguments suggesting its existence were given in
\cite{Witten,Hull}. Using this correspondence it is possible to
obtain information about Euclidean conformal field theories from
quantum gravity in de Sitter spacetime\cite{Recent}. It is significant that it does not seems possible to
formulate a unitary CFT for the massive fields $m^2 \ge
(\frac{d}{2}H)^2$, suggesting that such fields may not arise in a
consistent de Sitter quantum gravity. We will extend Strominger's analysis
to
the massless case and show that such fields do have a unitary CFT
interpretation. 

\section{De Sitter space and the representation theory}

The scalar representations of the de Sitter group $SO(1,d+1)$ split into three
series\cite{Vilenkin,Takook}, the principal series $m^2  \ge
(\frac{d}{2}H)^2$, the complementary series $0 < m^2 <
(\frac{d}{2}H)^2$ and the discrete series whose only physical
interesting case is $m^2=0$. Under a Wigner-In\"{o}n\"{u} contraction to the
Poincare group only the principal series representations contract to
Poincare representations and have an interpretation as particles in
the limit of zero cosmological constant\cite{Takook}. Strominger \cite{Strominger}
has shown that complementary series representations have a dual
interpretation in terms of a unitary CFT while the principal series
representations require a non-unitary description. However, as yet
the special case $m^2=0$ has not been treated. In the following we
shall concentrate on 4 dimensional de Sitter spacetime although
everything can easily be generalized to arbitrary dimensions. It is
important to consider the {\it de Sitter invariant} vacuum in order to
obtain information about the conformally invariant vacuum of the dual
CFT, and since its structure is subtle we shall consider in detail the
role of de Sitter invariance.

\smallskip 

The choice of vacuum for a massive scalar field in an arbitrary
spacetime is determined by the choice of complex structure
\cite{Wald}. This provides a decomposition of the space of real
solutions $\mathcal{S}$ to the equation of motion
$[-\nabla^2+m^2]\phi=0$ into two complex subspaces
$\mathcal{S}=\mathcal{V}^+ \oplus \mathcal{V}^- $ such that
$\mathcal{V}^+/\mathcal{V}^-$ has positive/negative Klein-Gordon norm
respectively. $\mathcal{V}^+$ is then taken to be the single-particle
Hilbert space $\mathcal{H}$ and the Hilbert space of the full theory
is then taken to be the usual Fock space
$\mathcal{F}=\mathcal{C}\oplus\mathcal{H}\oplus \mathcal{H} \otimes
\mathcal{H}\oplus \dots$. Demanding that the complex structure (and hence the
vacuum) be de Sitter invariant picks out a one-parameter family of
inequivalent de Sitter invariant vacua. Requiring that the short
distance behaviour of the Hadamard function should be of the same type as
that in flat spacetime then uniquely fixes the vacuum. This is known
as the Bunch-Davies or Euclidean vacuum\cite{Birrell}. In the massless
case the Klein-Gordon inner product and equation of motion are
invariant under the symmetry $\phi \rightarrow \phi +q$. Allen and
Folacci \cite{Allen} have
shown that this implies that there is no de Sitter invariant complex
structure and hence Fock space/vacuum. The situation is analogous to
QED and the resolution is to define a vacuum 
similar to the Gupta-Bleuler vacuum in
QED\cite{Takook} for which de Sitter invariance is
recovered. Consequently we need to look for a single-particle Hilbert
space defined as a space of positive norm solutions of the equations
of motion modulo symmetry transformations.

\smallskip

4d de Sitter spacetime is topologically $R \times S^3$ and can be covered with the metric
\begin{equation}
	ds^2=H^{-2}[-d \tau^2+\cosh^2\tau(d \chi^2 +\sin^2\chi(d
	\theta^2+\sin^2\theta d\phi^2))]
\end{equation}
with $-\infty < \tau < \infty$, $0 \le \chi, \theta \le \pi$,
$0 \le \phi \le 2 \pi$. These co-ordinates are known as the
closed slicing as they provide a foliation in terms of compact $S^3$
hypersurfaces. An arbitrary solution of the equations of motion can
then be written as a superposition over $S^3$ eigenmodes. Thus we are
lead naturally to the decomposition $\phi=\phi_g+\phi_z+\phi^++\phi^-$
\begin{eqnarray}
\nonumber & &
	\phi_g=const.	\\ & &
	\phi_z=H a_z(2\tan^{-1}\tanh(\tau/2)+\tanh\tau \> {\rm sech} \tau) \\ \nonumber & &
	\phi^+=	\Sigma_{L > 0, lm} a_{Llm} \chi_L(\tau) Y_{Llm}(\chi,\theta,\phi)\\ \nonumber & &
	\phi^-= \Sigma_{L > 0, lm} a^*_{Llm} \chi^*_L(\tau) Y^*_{Llm}(\chi,\theta,\phi)
\end{eqnarray} 
The $Y_{Llm}(\chi,\theta,\phi)$ are the
usual spherical harmonics defined on a 3-sphere. The $\chi_L(\tau)$ are defined by the
massless limit of the usual Bunch-Davies modes\cite{Birrell}
\begin{equation}
	\chi_L(\tau)=\frac{H}{2}A_L[L (\cosh \tau)^{L+2}(1+i\sinh \tau)^{-L-2}+
	(L+2) (\cosh \tau)^{L}(1+i\sinh \tau)^{-L})]
\end{equation}
with $A_L=[2 (L+2)(L+1)L]^{-1/2}$. Note that they satisfy the
reflection symmetry $\chi_L(\tau)^*=\chi_L(-\tau)$. We will denote the
subspaces of solutions by $\phi^+ \in \mathcal{V}^+$, $\phi^- \in \mathcal{V}^-$,
$\phi_g \in \mathcal{N}$, $\phi_z \in \mathcal{Z}$ so that the full
space of solutions is given by $\mathcal{S}=\mathcal{V}^+ \oplus
\mathcal{V}^- \oplus \mathcal{N} \oplus \mathcal{Z}$. The massless minimally coupled scalar field in de Sitter spacetime
belongs to the so-called discrete series representation of the de Sitter
group $SO(1,4)$ \cite{Vilenkin}. A candidate Hilbert space must have positive norm and
be closed under the action of the de Sitter group $T(g):\mathcal{H}
\rightarrow \mathcal{H}$. In addition the chosen norm must be
invariant, $\langle T(g)f,T(g)h\rangle=\langle f,h\rangle $. 
The natural inner product is the
Klein-Gordon inner product (see appendix). 
Under the action of an arbitrary element of the de Sitter group $g$ the subspaces
$\mathcal{V}^+$, $\mathcal{V}^-$, $\mathcal{N}$, $\mathcal{Z}$ transform as\cite{Takook}
\begin{eqnarray}
\nonumber & &
	T(g):{\mathcal{N}} \rightarrow {\mathcal{N}} \\ \nonumber & &
	T(g):{\mathcal{Z}} \rightarrow {\mathcal{Z}} \oplus
	{\mathcal{V^+}} \oplus {\mathcal{V^-}} 
	\oplus {\mathcal{N}} \\ \nonumber & &
	T(g):{\mathcal{V^+}} \rightarrow {\mathcal{V^+}} \oplus
	{\mathcal{N}} \\ \nonumber & &
	T(g):{\mathcal{V^-}} \rightarrow {\mathcal{V^-}} \oplus
	{\mathcal{N}}	\\ & &
\end{eqnarray}
$T(g)$ denotes the representation of the group element $g$ acting on
the space of solutions. To illustrate this we may construct the
$SO(1,4)$ generators with the help of
the embedding co-ordinates as
\begin{equation}
	M_{AB}=i[x_A \frac{\partial}{\partial x^B}-x_B \frac{\partial}{\partial x^A}]
\end{equation}
Then if we construct the following operator invariant under the $SO(4)$
subgroup (see appendix)
\begin{equation}
	\hat{Z}=i\frac{x^i}{\sqrt{x_i x^i}}M_{0i}=
	\frac{\partial}{\partial \tau}
\end{equation}
This operator represents the part of the generators that generates
time translations alone, it is only this part that mixes states of
different $L$.
One observes that
\begin{eqnarray}
\nonumber & &
	\hat{Z} \chi_1(\tau)=-\frac{i \sqrt{3}H}{8}- \frac{3i}{2}
	\chi_2(\tau) \\ & &
	\hat{Z} \phi_z(\tau)=a_z \sqrt{3} (\chi_1(\tau)+\chi_1(\tau)^*)
	\\ \nonumber & &
\end{eqnarray}
with the structure of the higher $L$ terms being always of the form $
\hat{Z} \chi_L(\tau)=a_L \chi_{L+1}(\tau)+b_L \chi_{L-1}(\tau)$.
It follows that the de Sitter invariant spaces are
${\mathcal{V^+}}/{\mathcal{N}}$ and
${\mathcal{V^-}}/{\mathcal{N}}$. Since the modes $\phi^+$ have
positive norm then the Hilbert space that forms the discrete series
representation of the de Sitter group is
$\mathcal{H}={\mathcal{V^+}}/{\mathcal{N}}$. Note that the
zero mode $\phi_z \in \mathcal{Z}$ does not contribute to
the physical state space since if it was included modes of positive
norm would be transformed into modes of negative norm violating
unitarity. In particular this means that only the $L > 0$ modes will
contribute to any physical quantity.
 
\section{Covariant quantization}

Covariant quantization begins with the fundamental quantization condition
\begin{equation}
	[\phi(x),\phi(x')]=\Delta(x,x') 
\end{equation}
where the commutator function
$\Delta(x,x')$ is equal to $i$ times the advanced minus retarded Green
functions. This incorporates causality from the start and ensures that
the commutator of two fields vanishes at spacelike separations. This
is distinct from the canonical approach where causality is not
immediately apparent and depends on which modes are included.
Note that $\Delta(x,x')$ is independent of the choice of vacuum state:
that information is contained in the Hadamard function we shall consider
later.

In the context of de Sitter spacetime 
$\Delta(x,x')$ has a smooth limit as the mass
$m$ of the field is taken to zero. So it may be defined as the
massless limit of the massive commutator\footnote{$\epsilon(x,x')$ is 1 ($-1$) if 
$x$ is in the future (past) of $x'$, and zero if $x$ and $x'$ 
are space-like separated.}
\begin{eqnarray}
\nonumber & &
	\Delta(x,x')=\epsilon(x,x')
	\frac{H^2}{8\pi^2}
	(\frac{1}{1-z+i \epsilon}-\ln|1-z+i \epsilon|
	-\frac{1}{1-z-i \epsilon}+\ln|1-z-i \epsilon|) \\ & &
	=-2 \pi i \epsilon(x,x')\frac{H^2}{8\pi^2}(\delta(1-z)+\theta(z-1))
\end{eqnarray}
where $z(x,x')$ is defined in the Appendix. We can then examine the
properties of 
$\Delta(x,x')$ in the closed coordinate system, by expanding it
in terms of modes on a 3-sphere.
Since $\Delta(x,x')$ is by definition a solution of
the equations of motion it must be a superposition of usual closed
slicing modes. In particular this includes the normalizable zero mode.
The result can be most usefully written in the form
\begin{equation}
[\phi(x),\phi(x')]=i\frac{H^2}{4 \pi^2}(f(\tau')-f(\tau))+W^+_n(x,x')-W^+_n(x',x)
\end{equation}
where $f(\tau)=2\tan^{-1} \tanh(\tau/2)+{\rm \tanh}\tau \> {\rm sech} \tau$ and can
be recognized as the $L=0$ mode term $\phi_z$.
Here $W^+_n(x,x')$ is the Wightman function constructed out of only the
$L > 0$ modes
\begin{eqnarray}
\nonumber & &
	W^+_n(x,x')=\Sigma_{L > 0 lm} \phi_{Llm}(x) \phi^*_{Llm}(x') \\
	& &
	=\frac{H^2}{8\pi^2}[\frac{1}{1-z}-\ln|1-z|-g(\tau)-g(\tau')+
	i(f(\tau)-f(\tau'))]+\frac{1}{2}\Delta(x,x')
\end{eqnarray}
where $g(\tau)={\rm sech}^2 \tau +\ln(2 {\rm sech} \tau)$. 
The expression for the field $\phi(x)$ which
gives rise to these commutation relations is
\begin{equation}
	\phi(x)=\frac{H}{2 \pi} Q+\frac{H}{2 \pi} f(\tau)P+\phi^+(x)+\phi^-(x)
\end{equation}
where $\phi^{\pm}$ is constructed entirely from modes $\in
\mathcal{V}^{\mp}$, with the commutation relations
\begin{eqnarray}
\nonumber & &
	{}[Q,P]=i, \\ & &
	[\phi^-(x),\phi^+(y)]=W^+_n(x,y),\\ \nonumber & &
	{}[\phi^+(x),\phi^+(y)]=[\phi^-(x),\phi^-(y)]=0.
\end{eqnarray}
Notice how the covariant quantization prescription has
determined the commutation relations between $Q$ and $P$.

The Hamiltonian (see section IV) is given by $\hat{H}=\frac{1}{2}(\cosh
\tau)^{-3}P^2+\hat{H}_{L > 0}$ where $\hat{H}_{L > 0}$ denotes the part of the Hamiltonian constructed
from the $L>0$ modes. Since $[P,\hat{H}]=0$, the operator $P$ is a conserved
generator that generates translations in $Q$. In particular we see that
\begin{equation}
	U(q)\phi(x)U^{\dag}(q)=\phi(x)+q
\end{equation}
where $U(q)=\exp[i \frac{2\pi}{H} Pq]$.
The commutation relations alone do not determine the correlators in the theory:
we need a prescription for picking the vacuum. In Minkowski
spacetime we define the vacuum to be the state annihilated 
by the positive energy  part of the field.  But since there
is no global timelike Killing vector in de Sitter spacetime, 
there is no global definition of positive energy, so the same
prescription cannot be applied.
There are three natural choices for the vacuum:

(i) The de Sitter invariant vacuum; 

(ii) The $\phi \rightarrow \phi+q$ invariant
vacuum\footnote{N.B. Invariance under $\phi \rightarrow \phi+q$ only tells us how to quantize
the $L=0$ mode};

(iii) The Euclidean vacuum.

To determine the de Sitter invariant state consider a single 
particle state generated by the action of the physical 
($\phi \rightarrow \phi+q$-invariant)
operator
 $\phi(x)-\phi(y)$
\begin{equation}
	|\psi>=(\phi(x)-\phi(y))|0>.
\end{equation}
We have already argued that if the single particle states are to form
a representation of the de Sitter group, the $L=0$ modes cannot
contribute. Since the $Q$ operator naturally cancels in constructing a
$\phi \rightarrow \phi+q$
invariant operator the $L=0$ contribution is linear in $P|0>$
and so de Sitter invariance demands $P|0>=0$. Alternatively we may see
that the requirement that 
action of a de Sitter group element on a single particle
state should not generate an $L=0$ mode implies 
\begin{equation}
	\int \hat{Z} \phi(x) d\Omega_{S^3} |0>=0
\end{equation}
We may also look for the state invariant under the
symmetry $\phi \rightarrow \phi+q$.  This implies that
\begin{equation}
	U(q)|0>=|0>
\end{equation}
and hence $P|0>=0$.
In what follows we refer to the vacuum defined by
$\{P|0>=0,\phi^-(x)|0>=0 \}$ as the Kirsten-Garriga (K-G) vacuum. We
shall show that this is equivalent to the Euclidean vacuum.

\section{Schr\"{o}dinger picture}

We begin by looking at the problem in the Schr\"{o}dinger
representation. The wavefunction $\Psi[\phi,\tau]$ satisfies the
Schr\"{o}dinger equation $i\frac{\partial \Psi}{\partial
\tau}=\hat{H}\Psi$ with time-dependent Hamiltonian
\begin{equation}
	\hat{H}(\tau)=H^{-1}\int d\Omega_{S^3}
	\frac{1}{2}[\frac{\pi^2}{(H^{-1} \cosh
	\tau)^{3}} +H^{-1} \cosh \tau
	(\nabla \phi)^2]
\end{equation}
where the canonical pair $\{\phi(x),\pi(x) \}$ satisfy the usual
canonical commutation relations on a 3-sphere. To elucidate we can
expand each of the fields in terms of modes on an $S^3$ with the
result that the Hamiltonian is given by
\begin{equation}
	\hat{H}=H^{-1} \Sigma_{Llm}
	\frac{1}{2}[\frac{\pi_{Llm}^2}{(H^{-1} \cosh
	\tau)^{3}} +H^{-1} \cosh \tau
	\omega_L^2 \phi_{Llm}^2 ]
\end{equation}
with $\omega_{L}^2=L(L+2)$. Following Kirsten and Garriga \cite{Garriga}, we may look for a gaussian
solution of the Schr\"{o}dinger equation of the form $\Psi = \Pi_{Llm} \Psi_{Llm}(\phi_{Llm},\tau)$
where
\begin{equation}
	\Psi_{Llm}(\phi_{Llm},\tau)=A_{Llm} \exp[\frac{iH}{2}(H^{-1} \cosh \tau)^{3}
	\frac{\dot{V}_{Llm}}{V_{Llm}} \phi_{Llm}^2]
\end{equation}
One can show that this is a solution to the Schr\"{o}dinger equation
providing the  $V_{Llm}$ satisfy
\begin{equation}
	\ddot{V}_{Llm}+3 \tanh \tau \dot{V}_{Llm} + {\rm sech}^2 \tau \omega_L^2 V_{Llm}=0
\end{equation}
along with $A_{Llm}=C_{Llm} V_{Llm}^{-1/2}$ and $C_{Llm}$ constant.
The above equation for the modes $V_{Llm}$ is just the classical equations of motion.  Until we specify which
set of solutions to this equation to take we do not have an explicit solution
of the Schr\"{o}dinger equation. Kirsten and Garriga go on to determine the
appropriate vacuum wavefunctional from the annihilation condition of
the wavefunction by the annihilation operators for the de Sitter
vacuum. 

\smallskip

The problem of the choice of vacuum becomes the
problem of what are the boundary conditions one should impose on
$\Psi$. One choice is that of the Euclidean 
no-boundary proposal \cite{Hawk2}, which by analogy with 
ordinary quantum mechanics (such as a harmonic oscillator) demands
that the wavefunction be regular when time is continued to 
imaginary values. In de Sitter spacetime, the most obvious continuation is
$\tau \rightarrow -iX$ which takes the de Sitter metric to that for
$S^4$. We must also 
analytically continue the solution of the Schr\"{o}dinger equation into
one half of the Euclidean $S^4$. This is achieved with $\tau
\rightarrow -iX$
\begin{equation}
	\Psi_{Llm}(\phi_{Llm},X)=A_{Llm} \exp[-\frac{H}{2}(H^{-1} \cos X)^{3}
	\frac{{V'}_{Llm}}{V_{Llm}} \phi_{Llm}^2]
\end{equation}
Here the $V_{Llm}$
satisfy the Euclidean equation of motion. We now demand that $\Psi$ is
real and is regular over bottom half of the $S^4$ ($0 \le X<\pi/2$). Now for $L \ge
0$ we require $\zeta = (\cos X)^{3} \frac{{V'}_{Llm}}{V_{Llm}}
\ge 0$ over the bottom half of the $S^4$ to ensure the convergence
of the wave-function normalization. The reflection symmetry of the
manifold means that the solutions are naturally split for $L
> 0$ into modes with $\zeta > 0$ and with $\zeta < 0$. In fact
the two sets of modes are, upon analytic
continuation, just the 
Bunch-Davies modes defined earlier with the
reflection property $\chi^*_L(-\tau)=\chi_L(\tau)$. So if
we choose $V_{L lm} = \chi_{L lm}$, we guarantee that the
exponent in 
(22) is negative, so the wavefunction is
normalizable.

\smallskip

For $L=0$ the most general solution is
$V_0=A+B(2\tanh^{-1}\tan(X/2)+{\rm sec} X \tan X)$ for which $\chi=2B/V_0$. 
The probability
$p(X)=\psi(-X)\psi(X)$ has an overall factor of
$(V_0(-X) V_0(X))^{-1/2}$. It is easy to see that for
nonzero $B$, $V_0(X)V_0(-X)$ vanishes somewhere in the Euclidean region
$0 \le X<\pi/2$.  Therefore if, as demanded in the no boundary prescription,
the wavefunction is to be regular in the Euclidean region, we must
have $B=0$
giving $\Psi_0=const$. In other words the
wavefunction is in a momentum eigenstate with $P=0$. Thus regularity
of the Euclidean wavefunction picks out the K-G vacuum.

\smallskip

Alternatively we may think in a completely covariant point of view
and consider the Euclidean Green's function equation
$\nabla^2 G(X,X')=\delta^4(X,X')$. As it stands this equation does not
make any sense since the $\phi \rightarrow \phi+q$ symmetry
means that $\nabla^2$ is not invertible. We can see this by
integrating over the 4-sphere, the left hand side vanishes as it is a
boundary term while the right is unity by definition. To make sense
of the formula one must 
project out the lowest $S^4$ spherical harmonic mode, i.e. the
constant mode, from the delta function\footnote{Not
to be confused with the $L=0$ mode previously discussed which is the
lowest $S^3$ spherical harmonic.}. The
solution for the Green's function is then
\smallskip
\begin{equation}
	G_E(Z)=\frac{H^2}{8 \pi^2} (\frac{1}{1-Z}-\ln(1-Z)+\ln2-\frac{1}{2}),
\end{equation}
which integrates to zero over the $S^4$.
Analytic continuation into the Lorentzian
manifold and symmetrization over $x$ and $x'$ gives the 
effective Hadamard function
\begin{equation}
	G^{h}(z)\equiv \langle\left\{\phi(x),\phi(x')\right\}\rangle
= \frac{H^2}{4 \pi^2}(\frac{1}{1-z}-\ln(1-z)+\ln2-\frac{1}{2})
\end{equation}
Note that this is not a solution of the equations of motion,
because we projected the constant mode out of the delta function.
But in fact $G^h$ differs from the symmetric two point function $\hat{G}(x,y)$
constructed from $L > 0$ modes only by the term
\begin{equation}
	G^{h}(x,x')-\hat{G}(x,x')=\frac{H^2}{4 \pi^2}(\ln2-\frac{1}{2}+g(\tau)+g(\tau')),
\end{equation}
and it is easy to see that
\begin{equation}
	\partial^{\mu}_x \partial^{\nu}_{x'} (G^{h}(x,x')-\hat{G}(x,x'))=0
\end{equation}
which implies that any correlators of {\it derivatives} of the scalar
field (from which all physical observables can be constructed) 
will be identical to those
obtained with only the $L \ge 0$ modes. In particular this applies
to the expectation value of the stress energy tensor, which
is defined by taking the limit as $x\rightarrow x'$, and performing a 
covariant subtraction.
Since the zero mode
contributes a term linear in $P^2$ to the stress energy it follows
that this stress-energy corresponds exactly to the value
calculated in the $P|0>=0$ vacuum\cite{Garriga}. In conclusion one may
define the theory in the Euclidean regime with the $S^4$ zero mode
projected and one obtains 
identical results for all physical
correlators as those obtained using the K-G vacuum.

\section{dS/CFT correspondence}

The de Sitter/CFT correspondence has been fueled by the hope that it
might make sense of quantum gravity in de Sitter spacetime. Motivated
by the fact that it does not seems possible to obtain de Sitter
spacetime as a solution of supergravity theories \cite{Maldacena} (for
recent work on evading these no-go theorems see \cite{dSsolutions} and
references therein), it
has been proposed that the Hilbert space describing quantum gravity
with a positive cosmological constant should be finite dimensional
\cite{Witten,Hull,Banks}. Witten has suggested that this problem might be
tackled by looking for a correspondence between quantum gravity on
asymptotically de Sitter spacetimes and a conformal field theory
living on the boundary of de Sitter spacetime. The conformal field
theory may then be used to define what we mean by quantum gravity with
a positive cosmological constant. As far as we are concerned here this
correspondence proposes a relation between the correlators of fields in de
Sitter spacetime and those of a CFT. Since this article appeared on
the archive a number of papers on the proposed de Sitter/CFT
correspondence have been published (see \cite{recent} and references therein).

\smallskip

In reference \cite{Strominger} a connection was found 
between scalar field theory correlators
in de Sitter and correlators of scalar operators in a dual
CFT. Ultimately this correspondence should be embedded into a
solution of M-theory in analogy with Maldacena's proposal of the
duality between large $N$, ${\mathcal{N}}=4$ super-Yang-Mills and low
energy $IIB$ string theory compactified on $AdS^5 \times
S^5$\cite{Maldacena1}. The low energy effective action of the
candidate solution may be taken to
be of the form
\begin{equation}
	S=\int d^{d+1}x \sqrt{-g} [ \frac{1}{2 \kappa} R
	-\frac{1}{2}G_{IJ}(\phi^K) 
	\partial \phi^I . \partial \phi^J -V(\phi^K)] 
\end{equation}
where $\phi^I$ denote a set of scalar fields and we have discarded
vector and fermionic fields. Non-minimal coupling to gravity that may arise through
the dilaton or from dimensional reduction has been absorbed by passing
to Einstein frame. The hope is to find a de
Sitter like solution $(g_0,\phi^K_0)$ for which $V'(\phi_0^K)=0$ and 
$V(\phi_0^K)>0$. In the region of this
minimum we may expand the potential as
$V(\phi^K)=V(\phi_0^K)+\frac{1}{2}m^2_{IJ} \phi^I \phi^J+\dots$ where we have
shifted the fields so that $\phi_0^K=0$. Thus if we consider
asymptotically de Sitter spacetimes the behaviour of the scalar fields
may be approximated by massive free fields. The effective masses $m_{IJ}$
are related to the renormalization group beta functions of the dual
field theory on the boundary that describe non-conformal effects. 
An exception to this occurs
if the potential has a 'flat direction' so that
$V(\phi^a+q)=V(\phi^a)$. In this case the effective mass in the
$\phi^a$ direction is zero and if it is also true that
$G_{IJ}(\phi^a+q)=G_{IJ}(\phi^a)$ the field $\phi^a$ will behave like
a massless minimally coupled scalar field in de Sitter spacetime.

\section{Massive case}

Following Strominger we consider a massive scalar field in de Sitter
spacetime using the flat co-ordinates $ds^2=H^{-2}(-d
t^2+e^{2t}(dx^2+dy^2+dz^2))$ (see Figure 1.). Future Infinity
$\tilde{I}^+$ lies at $t \rightarrow \infty$ and $t \rightarrow
-\infty$ is the horizon. The equation of motion has the
form
\begin{equation}
	-{\partial}^2_t \phi-3{\partial}_t \phi+e^{-2t} \nabla^2
	\phi=m^2 H^{-2} \phi
\end{equation}
The asymptotic behavior of the solutions at future
infinity\footnote{Strominger considers a contracting de Sitter
spacetime and so looks at the asymptotic behaviour at `past' infinity.} 
is $\phi \rightarrow e^{-h_{\pm} t}$
where $h_{\pm}=\frac{3}{2} \pm \sqrt{\frac{9}{4}-m^2H^{-2}}$. 
The Hadamard function for the massive field is given by 
\begin{equation}
	G^h(z)=\frac{H^2\Gamma[a_+] \Gamma[a_-]}{8 \pi^2}
{}_2F_1[a_+,a_-;2;\frac{1}{2}(1+z)]
\end{equation}
where $a_{\pm}=\frac{3}{2} \pm
(\frac{9}{4}-\frac{m^2}{H^2})^{1/2}$. Given a general solution with
asymptotic form
$\phi(x,t)=e^{-h_-t}\phi_-(x)+e^{-h_+t}\phi_+(x)$ the two point
correlator of the CFT can be obtained from considering the amplitude
\begin{equation}
	(\phi,\phi)=(\frac{i}{H})^2\int d^3x\int d^3x' e^{3(t+t')} \phi(t,x) {\mathop{\partial_t}\limits^\leftrightarrow}
	G^h(t,x;t',x') {\mathop{\partial_{t'}}\limits^\leftrightarrow} \phi(t',x')
\end{equation}
where $G^h(t,x;t',x')$ is the Hadamard function. The choice of Cauchy
surfaces is arbitrary but by taking them to future infinity one can
show that the above amplitude is given by

\begin{equation}
	c_- \int d^3x\int d^3x' \phi_-(x)|x-x'|^{-2h_+}\phi_-(x') +
	c_+ \int d^3x\int d^3x' \phi_+(x)|x-x'|^{-2h_-}\phi_+(x')
\end{equation}
where $c_{\pm}$ are constants. This shows that the $\phi_-$ modes are dual to a conformal field
theory operator $O^+(x)$ of dimension $h_+$ while the $\phi_+$ modes are dual
to an operator $O^-(x)$ of dimension $h_-$, so that
$<O^{\pm}(x)O^{\pm}(x')>=|x-x'|^{-2h_{\pm}}$.

\smallskip

An important difference between dS/CFT and AdS/CFT is
that in the former there are no 
non-normalizable modes in the bulk. In the AdS/CFT
picture it is the non-normalizable modes in the bulk that are dual to
sources on the boundary whilst normalizable modes are dual to
fields. In the dS/CFT case there are only normalizable modes giving
rise to two conformal fields on the boundary of different
dimension. Whether both fields are included depends on the choice of
asymptotic boundary conditions. We shall consider here the most
general form of boundary conditions.

\begin{figure}
\centerline{\psfig{file=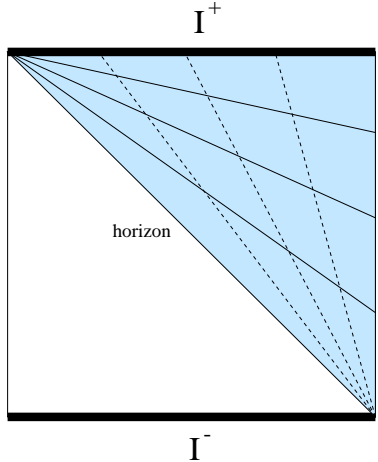,width=2.in}}
\caption{Penrose diagram for de Sitter spacetime: Future (Past)
Infinity $I^+$($I^-$) are 3-spheres. The shaded region is that covered
by the flat slicing bounded by the horizon of an observer at the
origin. Such an observer only sees future infinity $\tilde{I}^+$ as an
$R^3$ ($S^3$ minus one point), and only one point at past infinity due
to the horizon. The solid lines are surfaces of constant $t$ and the
dashed lines are surfaces of constant radius from the origin in the
flat slicing.}
\end{figure}

\section{Massless case - flat slicing}

The above reasoning does not apply in the $m \rightarrow 0$ limit. As
we shall see the asymptotic behaviour in this limit
changes. Furthermore the Hadamard function is infinite in the de
Sitter vacuum. Despite these problems one can still make sense of this
correspondence. A real
solution of the equations of motion may be expressed in a complete set
of modes
\begin{equation}
	\phi(x)=\int \frac{d^3k}{(2 \pi)^3} (a_k u_k(x)+a^{\dagger}_k u^*_k(x))
\end{equation}
with $u_k(x)=\frac{H}{\sqrt{2k}}(\eta-\frac{i}{k})e^{-ik\eta+ik.x}$
with $\eta=-e^{-t}$ where $-\infty < \eta < 0$.
Then one can show that above amplitude is given by
\begin{equation}
	(\phi,\phi)=2\int \frac{d^3k}{(2 \pi)^3} |a_k|^2
\end{equation}
Now since the flat slicing modes are well defined in the massless
limit then the amplitude is still well defined. The asymptotic
behaviour of the modes at future infinity is given by
\begin{equation}
	\phi(x,t)=\phi_-(x)-\frac{1}{2} e^{-2t} \nabla^2 \phi_-(x)+\phi_+(x) e^{-3t}+O(e^{-4t})
\end{equation}
It then remains only to relate $a_k$ in terms of
$\tilde{\phi}_{\pm}(k)$, the Fourier transform of $\phi_{\pm}(x)$. The result is 
\begin{equation}
	a_k=\frac{ik\sqrt{k}}{\sqrt{2}H}
	\tilde{\phi}_-(k)+\frac{3\sqrt{k}}{\sqrt{2} Hk^2}\tilde{\phi}_+(k)
\end{equation}
and so $(\phi,\phi)$ is given by
\begin{equation}
	\int d^3x\int d^3x' \phi_-(x)D_+(x,x')\phi_-(x') +
	\int d^3x\int d^3x' \phi_+(x) D_-(x,x') \phi_+(x')
\end{equation}
where
\begin{eqnarray}
\nonumber & &
	D_+(x,x')=\frac{1}{H^2}\int \frac{d^3k}{(2 \pi)^3}
	e^{ik.(x-x')}k^{3}=\frac{5!}{2H^2\pi^2}|x-x'|^{-6} \\  & &
	D_-(x,x')=\frac{9}{H^2}\int \frac{d^3k}{(2 \pi)^3} e^{ik.(x-x')}k^{-3}=c-\frac{9}{H^2\pi^2}\ln|x-x'|
\end{eqnarray}
The second correlator $D_-(x,x')$ exhibits the same infrared
divergence as the de Sitter Wightman function, and we have chosen to
separate out an infinite constant $c$ to see the physical behaviour. A
similar behaviour occurs for a massless field in two dimensions, for
instance as occurs in string theory. In that case the divergence comes
from the symmetry $X^{\mu} \rightarrow X^{\mu}+a^{\mu}$, i.e.
translation invariance on the target space. The result is that
$X^{\mu}$ is not a primary operator but its derivatives
$\partial X^{\mu}$ are. In
the Euclidean formulation where the worldsheet is taken to be an $S^2$
the divergence comes entirely from the $L=0$ $S^2$ spherical harmonic
term and can be projected out, the result is a logarithmic correlation
function. In view of these similarities we see that the holographic
dual of a massless minimally coupled scalar field in $dS^{d+1}$
contains two conformal fields $O^+(x)$ and $O^-(x)$, one primary with
conformal weight $d$ and correlator $<O^+(x)O^+(y)>=|x-x'|^{-2d}$ and a
second non-primary field $O^-(x)$, having global symmetry $O^-(x) \rightarrow
O^-(x)+a$ and correlator $<O^-(x)O^-(y)>=\ln|x-x'|$ whose
derivatives are primary fields. Alternatively we may in analogy with
string theory think in terms of the primary field $\exp[iq O^+(x)]$
whose correlator is given by
\begin{equation}
	<\exp[iq_1 O^-(x)] \exp[-iq_2 O^-(x)]>=\exp[-\frac{1}{2} q_1 q_2 <O^-(x)O^-(y)>]=|x-x'|^{-\frac{1}{2} q_1 q_2}
\end{equation}
so that $\exp[iq O^-(x)]$ has conformal dimension $q^2/4$.

\section{Massless case - closed slicing}

In the closed slicing one is free to use either future or past
infinity as the boundary. It is unnecessary to use both since the
fields at future infinity are completely determined by specifying
$\phi^-$ and $\phi^+$ at past infinity. It is for this reason that
only one copy of a CFT is needed. This reflects the fact that the
holographic dual of the bulk lives not so much on the boundary but
rather on the effective Cauchy surface. We shall define the
holographic dual at future infinity for ease of comparison with the
flat slicing case.
In order to give a finite amplitude we choose a $\phi(x)$ such that
\begin{equation}
	\int \frac{\partial \phi(x)}{\partial \tau} d\Omega_{S^3}=0
\end{equation}
It is significant that the $L=0$ mode has not been included. This
allows us to work with manifestly finite expressions. The asymptotic behaviour
at future infinity is
\begin{equation}
	\phi(x,\tau)=\phi_-(x)-2e^{-2\tau} \nabla^2 \phi_-(x)+\phi_+(x) e^{-3\tau}+O(e^{-4\tau})
\end{equation}
Ignoring the contributions form the $L=0$ modes the asymptotic form of
the Hadamard function is
\begin{equation}
	G^{h}(x,\tau;x',\tau')=\frac{H^2}{4\pi^2}(-\ln \sigma+(1-\frac{2}{\sigma})(e^{-2\tau}+e^{-2\tau'})-4(\frac{1}{\sigma^2}+\frac{1}{\sigma})e^{-2(\tau+\tau')}+\frac{128}{3\sigma^3}e^{-3(\tau+\tau')})
\end{equation}
to order $O(e^{-4\tau},e^{-4\tau'})$. Then 
\begin{equation}
	(\phi,\phi)
	=c_+ \int d\Omega_{S^3} \int d\Omega'_{S^3}
	\phi_-(x) \frac{1}{\sigma^3} \phi_-(x')+c_- \int d\Omega_{S^3} \int d\Omega'_{S^3}
	\phi_+(x) \ln \sigma \phi_+(x')
\end{equation}
where $c_{\pm}$ are constants. Note that the infinite constant that
occurred in the flat slicing correlator has been automatically
projected out by imposing the condition (e.q. 16).
As expected this is simply the flat slicing result conformally
transformed to the 3-sphere. 

\section{Conclusion}

In this paper we have shown how to consistently quantize the massless
minimally coupled scalar field on de Sitter spacetime for the de
Sitter invariant vacuum. We have shown that this is equivalent to the
Euclidean vacuum as for the massive case. This treatment shows the
different but nevertheless equivalent ways global symmetries are
treated in Lorentzian and Euclidean formulations of field theory. We
have shown that as for the massive case the massless minimally coupled
scalar field also has a CFT interpretation as a conformal field with weight
$d$ and a second field with a logarithmic correlator. The second field
exhibits the same global symmetry and infrared divergence as the
massless field in de Sitter and we argue that this apparent problem
should have the same resolution, namely that one should think in terms
of physical correlators for which the infrared divergence is
removed. It is well known that the graviton in de Sitter spacetime
considered in physical gauge in the flat slicing behaves like a
massless minimally coupled scalar field \cite{Woodard} and so we
anticipate that a similar approach will be needed in this case.

\smallskip
\
{\bf Acknowledgements:} 
We thank C. Galfard and
T. Wiseman for useful discussions. AJT acknowledges the
support of an EPSRC studentship. The work of NT is supported by
PPARC (UK).

\section{Appendix}

De Sitter spacetime can be
defined as the hyperboloid $x\cdot x=H^{-2}$ embedded in a 5 dimensional flat
manifold with signature $(-++++)$, where `$\cdot$' 
denotes the $SO(1,4)$ invariant inner product. The constant
`$H$' is the Hubble constant viewed from the flat slicing. The 
curvature is given by $R=12H^2$.
The distance between two points in the embedding space $(x-y)\cdot(x-y)$
provides a de Sitter invariant metric.
It is usual to
define the quadratic form $z(x,y)$ through $\frac{H^2}{2}(x-y)^2=(1-z(x,y))$,
so that 
$z(x,y)=H^2 x\cdot y$.
If $z>1$ ($<1$) the points are timelike (spacelike) separated. 

The closed co-ordinate system is obtained through the embedding
procedure by the identification
$(x^0,x^i)=(H^{-1} \sinh \tau,H^{-1} n^i \cosh \tau) $
where $n^i n_i=1$ defines an $S^3$. Explicitly 
$n^i=(\sin \chi \sin \theta \cos \phi, \sin \chi \sin \theta \sin
\phi,\sin \chi \cos \theta, \cos \chi)$.
The de Sitter invariant measure, or volume element, in this
co-ordinate system is given by
\begin{equation}
	dV=\sqrt{-g}d^4x=H^{-4}(\cosh \tau)^{3} d\tau d\Omega_{S^3}
\end{equation}
where $d \Omega_{S^3} =\sin^2 \chi \sin \theta d\chi d\theta d\phi$ is the
volume element on
$S^3$. The invariant distance between two points
on $S^3$, used in the text, is defined by
$\sigma(x,x')=\frac{1}{2}|n^i-{n^i}'|^2=1-n^i n_i'$.

The Klein-Gordon inner product is given by
\begin{equation}
	\langle \phi,\psi \rangle =\frac{i}{H^2}\int_{\Sigma} (\cosh \tau)^3 (\phi^*
	\partial_{\tau} \psi-\psi^* \partial_{\tau} \phi) d\Omega_{S^3}
\end{equation}
In the scalar representation of $SO(1,4)$, the generators are
\begin{equation}
	M_{AB}=i[x_A \frac{\partial}{\partial x^B}-x_B \frac{\partial}{\partial x^A}].
\end{equation}
In addition to the 
$SO(4)$ generators $M_{ij}$ associated with the isometry group of the
$S^3$, one has the non-compact generators 
\begin{equation}
	M_{0i}=-in_i \frac{\partial}{\partial \tau}-i\tanh \tau[
	\frac{\partial}{\partial n^i}-n^i n^j\frac{\partial}{\partial n^j}],
\end{equation}
and so
\begin{equation}
	\hat{Z}=i\frac{x^i}{\sqrt{x^ix_i}} M_{0i}=in^iM_{0i}=\frac{\partial}{\partial \tau}.
\end{equation}

The analytic continuation from de Sitter spacetime to $S^4$ is achieved
by continuing 
$\tau \rightarrow -iX$ on the matching surface $\tau=X=0$.
This yields the Euclidean metric 
\begin{equation}
	ds^2=H^{-2}(dX^2+\cos^2 X (d\chi^2+\sin^2 \chi (d\theta^2+\sin^2 \theta d \phi^2)))
\end{equation}
with $-\pi/2 < X < \pi/2$. Distances on the $S^4$ are defined 
via $Z(x,y)=H^2
x\cdot y$, just as in the de Sitter case but now using the
$O(5)$ invariant metric on the flat embedding space.

\def\Journal#1#2#3#4{{#1}{\bf #2}, #3 (#4)}
\def\NPB{Nucl.\ Phys.\ {\bf B}}
\def\PLB{Phys.\ Lett.\ {\bf B}}
\def\PRL{Phys.\ Rev.\ Lett.\ }
\def\PRD{Phys.\ Rev.\ D }
\def\JPA{J.\ Phys.\ {\bf A}}
\def\CMP{Commun.\ Math.\ Phys.\ {\bf A}}
\def\IJMP{Int.\ J.\ Mod.\ Phys.\ {\bf A}}
\def\JHEP{J.\ High Energy Phys. }

\end{document}